# Electronic and Structural Analysis of a Stable Hydrogenated BN sheet (BHNH): A First Principles Based Approach

AUTHOR NAMES: A. Bhattacharya[1], S. Bhattacharya[1], C. Majumdar[2], G.P. Das[1*]

AUTHOR ADDRESS: [1]Department of Materials Science, Indian Association for the Cultivation of Science, Jadavpur, Kolkata 700032, India,

[2]Chemistry Division, Bhabha Atomic Research Centre, Mumbai 400085, India,

[1*]Department of Materials Science, Indian Association for the Cultivation of Science, Jadavpur, Kolkata. 700032, India.

AUTHOR EMAIL ADDRESS: [1]msab2@iacs.res.in, [1]mssb3@iacs.res.in, [2]chimaju@barc.gov.in, [1*]msgpd@iacs.res.in.

CORRESPONDING AUTHOR FOOTNOTE: [1*]G. P. Das, Department of Materials Science, Indian Association for the Cultivation of Science, Jadavpur, Kolkata-700032, India. Phone No: +91-33-2473 4971 Extn. 202 (Office), Fax: +91-33-2473 2805.

ABSTRACT: From first-principles density functional calculations, we study the structural and electronic properties of a stable hydrogenated BN sheet, having formula unit BHNH. In the optimized BHNH structure, the H atoms stabilize on the B and N sites, alternating themselves on both sides of the BN-plane in specific periodic manner, giving rise to different BHNH conformers, viz. chair, boat and stirrup. The chair and boat conformers resemble in structure to those of graphane (CH). We propose a new conformer, called 'stirrup' conformer, that turns out to be the most stable, albeit marginally with respect to the boat conformer. All these BHNH conformers are insulator, with band gap varying between ~ 3.0 eV to 4.5 eV.

MANUSCRIPT TEXT:

Introduction:

After the discovery of graphene, which is a two dimensional (2D) array of hexagonal units of $sp^2$-bonded C atoms[1], there has been an avalanche of research activities both from the point of view of fundamental understanding as well as application potential of this wonder-sheet. Recently, a hydrogenated-graphene sheet called 'graphane' having a formula unit CH was reported[2] and has been found to be a highly stable insulating structure that can exist as a single layer. Subsequently, graphane was synthesized in laboratory by Geim's group[3] at Manchester University. Graphane is reported to have two favorable conformations *viz*. 'chair' and 'boat'. In both the conformers all C atoms are bonded to one H atom each, but in different orientations. In the 'chair' form of graphane, H atoms alternate on both sides of the plane, while in the 'boat' form the H atoms alternates in pair on both sides of the plane. Graphane could have prospected application in low dimensional electronics involving high mobility electron gas with variable electron concentration.

An analogous 2D atomic crystal *viz*. BN sheet has been cited as a possible analogue of graphene in various ways[4-6]. BN has various allotropic modifications that are known to possess unique structural and electronic properties. Pairs of B and N atoms in BN sheet make it iso-electronic with pairs of C atoms in graphene. BN sheet has been experimentally synthesized in single and multiple layers[7]. Although both have similar crystal structure with very close cell parameters, their nature of bonding and electronic properties are remarkably distinct. Graphene can be metallic, semi-metallic or semiconducting, depending upon its edge structure.[8] On the contrary; the *h*-BN sheet is typically a wide band gap (~4.7eV) insulator[7].

It is worth looking into the possibility of hydrogenation of BN sheet, in the same spirit as graphane, and continuing study on its structural and electronic properties. In recent past their have been various reports on hydrogenated BN sheet.[9] The objective of this letter is to investigate the relative stability of the different conformers of hydrogenated BN sheet (BHNH sheet) having the chemical formula unit BHNH. In the BHNH sheet, each of the B and N atoms are bonded to one H atom, and in some specific manners that give rise to various conformations of the sheet. These conformers show reasonably high stability (as will be discussed below), and can exist in single layer (with negligible or low interlayer binding) whose structural, electronic and vibrational properties have been calculated. As in the case of graphane, here also we have the possibility of having 'chair' and 'boat' conformers[2] along with a new 'stirrup' conformer (Fig. 1).

Our calculations have been carried out using first-principles density functional theory (DFT)[10,11] based total energy calculations with generalized gradient approximation (GGA). We have used VASP[12] code with projected augmented wave (PAW) potential[13] used for all elemental constituents; viz. H, Li, B and N. The GGA calculations have been performed using the exchange correlation functional of Perdew et al[14,15]. An energy cut off of 600eV has been used. The K-mesh [8 X 8 X 1] was generated by Monkhorst–Pack[16] method, and all results were tested for convergence with respect to mesh size. In all our calculations, self-consistency has been achieved with a 0.0001 eV convergence in total energy. For optimizing ground state geometry[17,18], atomic forces were converged to less than 0.001eV/Å via conjugate gradient minimization. We have implemented the Hirshfeld charge analysis scheme to obtain the detailed charge transfer mechanism in all atoms comprising the system. A pure BN sheet has two atoms (1B and 1N) as basis in one primitive cell. The simulation cell was modeled by taking a 4X4 triclinic super cell of dimension 9.8396 Å X 9.8396 Å X 16 Å, containing 16 B and 16 N atoms in it.

**Stability analysis:** BN sheet can be looked upon as 2D array of repetitive hexagonal BN units where each B (N) atom is bonded to 3 N (B) atoms.[19] It has high stability with an estimated binding energy of ~ 7 eV/atom. It is an insulator with a band-gap of 4.7eV. Hirshfeld's charge analysis shows that each B atom in the sheet has a charge state of +0.19 while that of N atom is -0.19. In this context, we would like to mention that borazine ($B_3N_3H_6$) and having the same formula unit as that of BHNH sheet, is a stable compound with BE of 4.45 eV/atom. It is a planar molecule having hexagonal BN unit connected to 6 H atoms. Unlike the case of borazine, where all atoms are in the same plane, each of the B and N atoms in BHNH sheet has only one H atom bonded to it in different orientations. From the energetic point of view, the stirrup BHNH conformer turns out to be the most stable one having a BE of 4.65 eV/atom which is higher than that of borazine by about 0.2eV/atom. All the conformers of BHNH sheet have a symmetrically puckered zigzag 3D structure. The boat and chair BHNH conformers resemble in structure to those of graphane[1]. However result shows that the 'stirrup' conformer of BHNH sheet has got a new structure that resembles the structure of cubic BN (100) surface.

**Structural properties of different BHNH conformers:** In table 1 we have enlisted various structural properties of these three conformers, some of which are discussed below.

**(a) Chair conformer**: In the chair BHNH conformer, consecutive H atoms alternate on both sides of the plane in a way that all the H atoms bonded to B atoms ($H_B$) are on one side of the plane, while all the H atoms bonded to N atoms ($H_N$) lie on the other side of the plane (fig 1a, b). The calculated bond lengths of B-N, B-H, and N-H bonds in chair BHNH conformer are 1.52 Å, 1.20 Å and 1.03 Å respectively (table 1). It has binding energy and band gap of 4.54 eV/ atom and ~ 2.84 eV respectively, which are the lowest amongst all the three conformers (fig 2). The interlayer binding between two layers of chair-BHNH conformer is found to be 0.02 eV/ atom, which is highest amongst all BHNH conformers. Hirshfeld charge analysis (table 2) shows that in a chair conformer, all the $H_B$ atoms (with charge -0.045) lie on one side of the plane, while all the $H_N$ atoms (with charge +0.092) lie on the other side of the plane (table 2). Thus, the oppositely charged $H_B$ and $H_N$ planes experience an attractive ionic interaction leading to the bonding between two consecutive layers of this conformer.

**(b) Boat conformer**: In the boat conformer, the H atoms alternate in pair on both sides of the plane (fig 1c,d ). Unlike in case (a), the B-N bonds in the boat conformer show two different bond lengths which results from the different H-environment of the B and N atoms, viz (i) 1.56 Å connecting B and N and bonded to hydrogen atoms in the same side of the plane and (ii) 1.51 Å connecting B and N bonded to hydrogen atoms on opposite side of the plane (table 1). The B-H and N-H bond distances are 1.12 Å and 1.01 Å respectively. Our calculation shows that the boat BHNH conformer has binding energy and band gap of 4.81 eV/atom and 4.23 eV (fig 2) respectively. However, the binding between two consecutive layers of boat conformer is very weak. This is because in boat conformer, the H-atoms of opposite polarity i.e., $H_B$ (-0.023) and $H_N$ (+0.240), exists in each plane. Thus, the adjacent layers facing each other experiences both attractive ($H_N$-$H_B$) and repulsive interactions ($H_B$-$H_B$ and $H_N$-$H_N$).

**(c) Stirrup conformer:** In stirrup BHNH conformer, three consecutive B and N atoms are connected to H-atoms all above the plane, while the next three consecutive B and N atoms have their H-atoms attached all below the plane (fig 1e,f). Out of the three conformers (viz. chair, boat and stirrup), the stirrup conformer exhibits the highest stability with a binding energy of ~4.85 eV/atom which is nearly degenerate to that of the boat conformer. Thus we observe a new conformer that has not been reported so far in the literature. It has two different B-N bond lengths, which are 1.61 Å and 1.57 Å (table1). The B-H and N-H bond length are 1.21 Å and 1.03 Å respectively. It is an insulator having the highest band gap of ~ 4.31 eV amongst the three BHNH conformers (fig 2). The interlayer bonding between two layers of BHNH stirrup conformer is also quite weak because of the same reason as mentioned in case (b). This is responsible for the pair-wise dimerization of the B-N bonds, as can be seen from Fig. 1(f).

**Vibrational frequency analysis:** The vibrational frequency analysis shows that all the three BHNH conformers have the highest normal mode vibrational frequency in the IR region of frequency spectrum and it corresponds to N-H stretching mode for all the conformers. The highest vibrational normal mode frequency of the chair and boat conformers is 3272.5 cm$^{-1}$ and 3265.4 cm$^{-1}$ respectively while the stirrup conformer records the highest normal mode vibrational frequency of 3345.8 cm$^{-1}$.

**Charge transfer**: In order to understand the charge transfer mechanism we have performed the Hirshfeld's charge analysis for all the three conformers. Let us denote the amount of electronic charge transferred in B by ΔB (difference in the charge state of B in pure BN sheet and that in BHNH conformer) and similarly the same for N by ΔN. The charge transfers (see Table 2) in chair and stirrup conformer show similar trend in the sense that $H_N$ and N get depleted of electrons, while those electrons are gained by B and $H_B$. Fig. 3 shows the channel in which the charge transfer takes place. In all the cases $H_N$ loses electron while $H_B$ gains electron and acquires a positive charge state. In chair BHNH conformer, the net amount of electronic charge gained by B atom is ΔB =-0.112, while that lost from N atom is ΔN = +0.065. In case of stirrup conformer, B atoms accept electron (ΔB =-0.097) and N atom loses electron (ΔN = +0.069) (table 2). In contrast, the charge transfer in boat conformer differs significantly from the other two. Hirshfeld's charge analysis for boat conformer shows that $H_N$ loses electron significantly (+0.24), and this is distributed amongst N, B and $H_B$ atoms. Only in this boat conformer both B and N gains electron with an amount given by ΔB = -0.068 and ΔN = -0.148 (table 2)

**Electronic structure analysis**: The salient features of the electronic structures of the three different conformers can be seen from the band structure in the Γ→M→K plane and the site-projected densities of states as shown in Fig. 2. While the occupied part of DOS remain more or less same (~10eV) with strong hybridization between B, N and H, the GGA band gaps increase from ~ 2.8-4.3eV as one goes from Chair to Stirrup conformer. All are direct band gap semiconductors with conduction band minimum and valence band maximum appearing at the Γ-point, while the band dispersion along M→K region is almost flat. The stirrup conformer characterized by the largest band gap and a deep pseudo-gap just below the valence band maximum is the most stable. It is to be noted here that partial DOS of boat and stirrup conformers are quite similar with nearly equal band gap and peaks at more or less same energies, thereby suggesting the possibility of existence of these two nearly degenerate conformers of BHNH sheet.

**Conclusion:** In summary, we report the ground state structure, stability and electronic properties of hydrogenated BN sheet on the basis of our first principles DFT based calculations. The geometry optimized structure shows BH and NH bonds alternating themselves on both side of the plane in specific periodic manner. There are three possible conformers viz. chair, boat and stirrup which are all insulating and having direct band gaps varying between 3eV and 4.5eV. The stirrup conformer turns out to be the most stable and its structure resembles to that of the puckered BN (100) surface.

FIGURE CAPTIONS (Color Online):

Fig 1: Ball and stick model of different BHNH conformers, with pink-balls representing Boron, blue-balls representing Nitrogen, and white-balls representing Hydrogen : (a) Chair conformer (c) Boat conformer (e) Stirrup conformer, and their corresponding layer views (b), (d) and (f) respectively.

Fig 2: Band structure and density of state (both Total and Partial) plots of (a) Chair conformer (b) Boat conformer (c) Stirrup conformer of BHNH sheet. All are direct band gap insulators, with increasing band gaps.

Fig 3: Charge transfer channel of BHNH conformers.

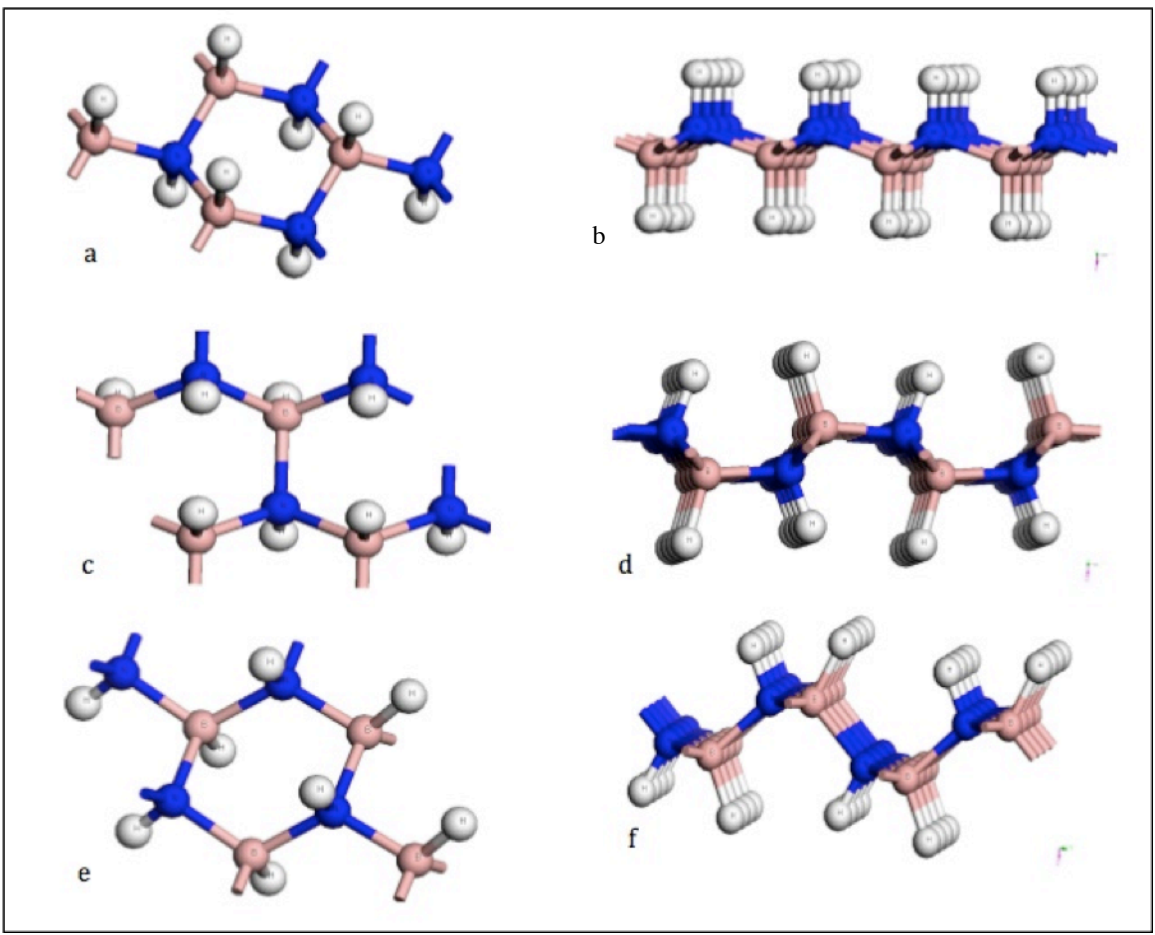

Fig 1.

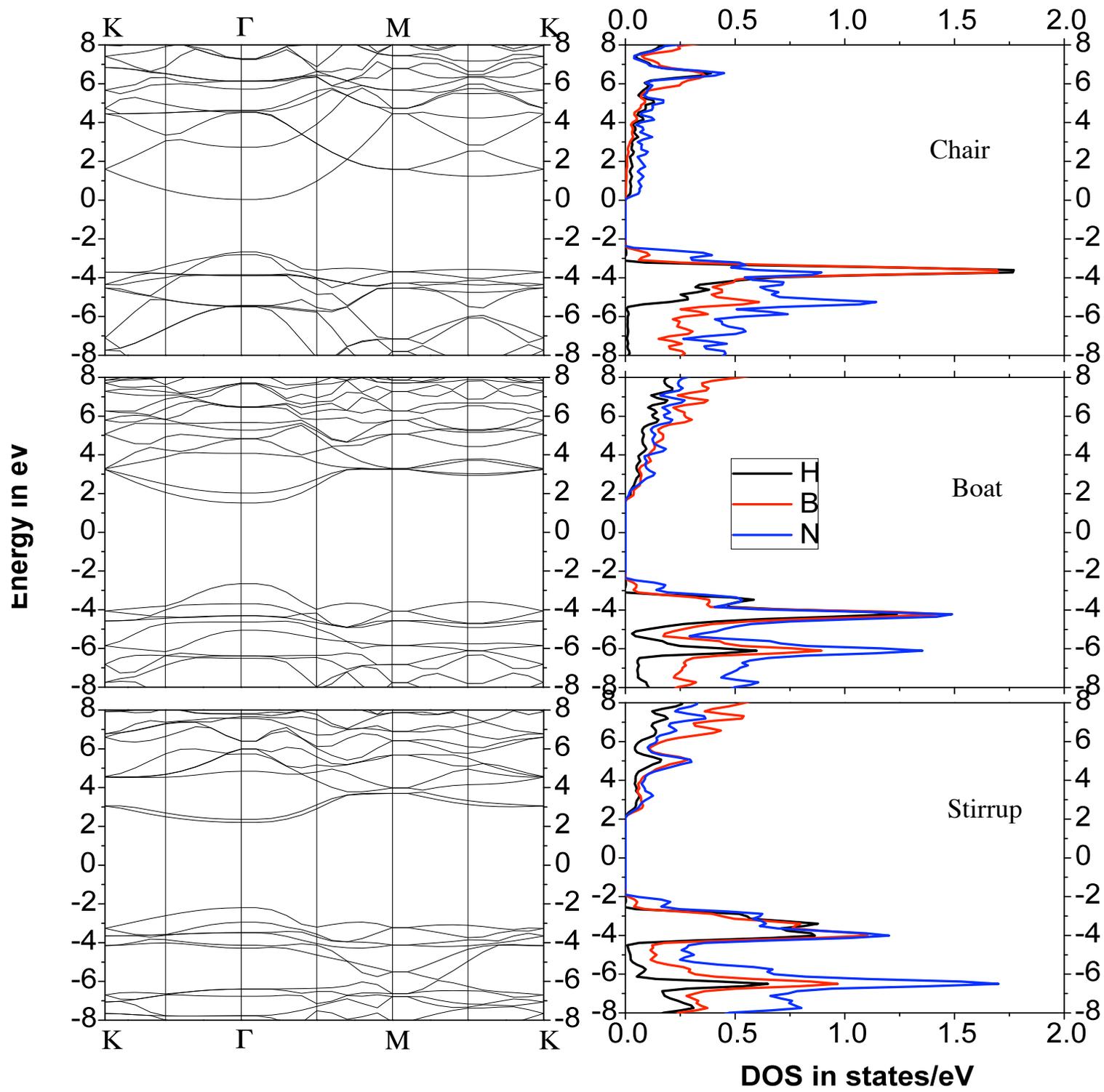

Fig 2.

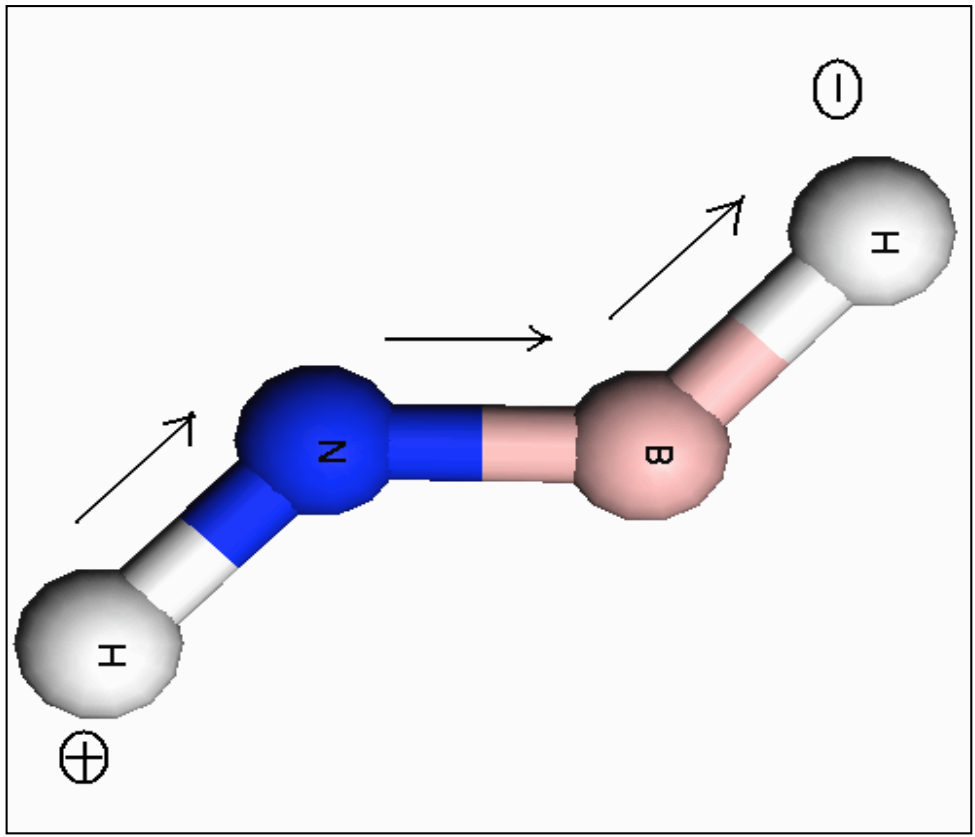

Fig: 3

TABLES.

Table1. Comparison of the various ground state properties of 'chair', 'boat' and 'stirrup' conformers of the BHNH sheet.

| Properties | Details | Chair | Boat | Stirrup |
|---|---|---|---|---|
| Bond length (Å) | B-H | 1.192 | 1.12 | 1.22 |
| | N-H | 1.03 | 1.01 | 1.03 |
| | B-N | 1.52 | 1.51,1.56 | 1.61,1.57 |
| Stability (eV/atom) | | 4.54 | 4.81 | 4.85 |
| Inter layer bonding (eV/atom) | | 0.02 | 0.005 | 0.004 |
| Energy gap (eV) | | 2.84 | 4.23 | 4.31 |
| Frequency of highest vibrational mode (cm$^{-1}$) | | 3272.5 | 3265.4 | 3345.8 |

Table 2. Hirshfeld charge analysis showing charge states of atoms. (Negative sign indicates electron gained by the atom while positive sign implies charge lost by the atom). ΔB and ΔN shows the magnitude of charge gain/lost from pure B and N atoms respectively.

| System | B | ΔB | N | ΔN | H$_B$ | H$_N$ |
|---|---|---|---|---|---|---|
| Pure BN | +0.194 | | -0.194 | | | |
| Chair BHNH | +0.082 | -0.112 | -0.129 | +0.065 | -0.044 | +0.092 |
| Boat BHNH | +0.126 | -0.068 | -0.342 | -0.148 | -0.023 | +0.239 |
| Stirrup BHNH | +0.098 | -0.096 | -0.125 | +0.069 | -0.085 | +0.118 |

REFERENCES

1. A. K. Geim and K. S. Novoselov Nat. Mater. **6**, 183 (2007)

2. J. O. Sofo, A. S. Chaudhari and G D. Barber, Phys. Rev. **75**, 153401 (2007).

3. D. C. Elias, R. R. Nair, T. M. G. Mohiuddin, S. V. Morozov, P. Blake, M. P. Halsall, A. C. Ferrari, D. W. Boukhvalov, M. I. Katsnelson, A. K. Geim and K. S. Novoselov, Science **323**, 610 (2009).

4. K. S. Novoselov, A. K. Geim, S. V. Morozov, D. Jiang, Y. Zhang, S. V. Dubonos, I. V. Grigorieva and A. A. Firsov, Science **22**, 306 (2004).

5. M. Terrones, J.M. Romo-Herrera, E. Cruz-Silva, F. López-Urías, E. Muñoz-Sandoval, J.J. Velázquez-Salazar, H. Terrones, Y. Bando and D. Golberg Materials Today **10**, 30 (2007).

6. C. Azevedo, A. Sadanandom, K. Kitagawa, A. Freialdenhoven, K. Shirasu and P. Schulze-Lefer Science **15**, 2073 (2002).

7. A. Nag, K. Raidongia, K. Hembram, R. Datta, U. V. Wagmare and C.N.R. Rao ACS Nano **4**, 1539 (2010); C. Jin, F. Lin, K. Suenaga and S. Iijima Phys. Rev. Lett. **102**, 195505 (2009).

8. V.B. Shenoy, C.D. Reddy, A. Ramasubramaniam, and Y.W. Zhang Phys. Rev. Lett. **101**, 245501 (2008).

9. J. Zhou, Q. Wang, Q. Sun and P. Jena Phys. Rev. B **81**, 085442 (2010); F.W. Averill, J.R. Morris and V.R. Cooper Phys. Rev. B 194411 (2009).

10. P. Hohenberg and W. Kohn Phys. Rev. **136**, B864 (1964).

11. W. Kohn and L. Sham Phys. Rev. **140**, A1133 (1965).

12. G. Kresse and J. Hafner, Phys. Rev. B **49**, 14251 (1994); G. Kresse and J. J. Furthmüller Comput. Mater. Sci. **6**, 15 (1996).

13. P. E. Blöchl Projector augmented-wave method. Phys. Rev. B **50**, 17953 (1994).

14. J. P. Perdew and Y. Wang Phys. Rev. B **45**, 13244 (1992).

15. J. P. Perdew, J.A. Chevary, S. H. Vosko, K. A. Jackson, M. R. Pederson, D. J. Singh and C. Fiolhais Phys. Rev. B **46**, 6671 (1992).

16. H. J. Monkhorst and J. D. Pack Phys. Rev. B **13**, 5188 (1976).

17. W. H. Press, B. P. Flannery, S. A. Tenkolsky and W. T. Vetterling Numerical Recipes (Cambridge University Press, New York, 1986), Vol. 1.

18. P. Pulay Chem. Phys. Lett. **73**, 393 (1980).

19. G. Giovannetti, P. A. Khomyakov, G. Brocks, P. J. Kelly and J. V. Brink, Phys. Rev. B **76**, 073103 (2007).